\documentclass[a4paper,10pt]{article}
\usepackage{amsmath}
\usepackage{amssymb}

\usepackage{graphicx}
\input epsf.tex
\usepackage{amsthm}

\newcommand{\be}{\begin{equation}}
\newcommand{\ee}{\end{equation}}
\newcommand{\ba}{\begin{eqnarray}}
\newcommand{\ea}{\end{eqnarray}}
\newcommand{\ban}{\begin{eqnarray*}}
\newcommand{\ean}{\end{eqnarray*}}

\newcommand*{\EC}{\mathrm{EC}}
\newcommand*{\PA}{\mathrm{PA}}
\newcommand*{\PE}{\mathrm{PE}}
\newcommand*{\leak}{\mathrm{leak}}
\newcommand*{\eps}{\varepsilon}

\newcommand{\demi}{\frac{1}{2}}

\newcommand{\one}{\leavevmode\hbox{\small1\normalsize\kern-.33em1}}

\begin{document}

\title{Finite-key analysis for practical implementations of quantum key distribution}
\author{Raymond Y.Q. Cai, Valerio Scarani\\ Centre for Quantum Technologies and Department of Physics\\National University of Singapore, Singapore}
\date{\today}
\maketitle

\begin{abstract}
The lists of bits processed in quantum key distribution are necessarily of finite length. The need for finite-key unconditional security bounds has been recognized long ago, but the theoretical tools have become available only very recently. We provide finite-key unconditional security bounds for two practical implementations of the Bennett-Brassard 1984 coding: prepare-and-measure implementations without decoy states, and entanglement-based implementations. A finite-key bound for prepare-and-measure implementations with decoy states is also derived under a simplified treatment of the statistical fluctuations. The presentation is tailored to allow direct application of the bounds in experiments. Finally, the bounds are also evaluated on \textit{a priori} reasonable expected values of the observed parameters.
\end{abstract}

\section{Introduction}

In 1984, Bennett and Brassard remarked that quantum physics provides a solution to the cryptographic task of distributing a secret key and provided the first explicit protocol, known as BB84 \cite{bb84}. This fact was re-discovered in 1991 by Ekert \cite{eke91}. Since, quantum key distribution (QKD) has grown into a mature field, spanning a wide range of competences; several reviews have been devoted to it \cite{gis02,dus06,sca08,lo08}.

The fast development of QKD can be tracked down to the interplay of two factors. First: QKD allows \textit{unconditional security} \cite{may96,LoChau,ShoPre00,Mayers01,BenOr02,KGR,rennerthesis,Koashi}, which means that security can be guaranteed in an information-theoretical sense, without any assumption on the computational power of the eavesdropper. Therefore, the task in itself is interesting, because it reaches beyond anything that can be done with classical communication alone. Second: QKD can be implemented without entanglement \cite{bb84} or with one entangled pair \cite{eke91} and has therefore been well within reach of existing experimental technologies for several decades.

The matching of a theoretical security proof to a real device is however a delicate matter. On the one hand, while unconditional security does not put any constraint on the eavesdropper, the proofs \textit{do} contain assumptions about the behavior of the devices of the authorized partners: the quantum states that are prepared, the model of the detectors, the procedures used for the classical post-processing of the data... On the other hand, imperfections of the real devices may leak information in side channels or allow for Trojan Horse attacks or other purely classical hacking attacks \cite{kur01,makarov,zhao07}: it is clearly impossible to devise a security proof that would take all these failures into account (for the so-called \textit{device-independent} approach to security and its assumptions, we refer to \cite{aci07,pir09}). The development of checking procedures based on testable assumptions is one of the most urgent tasks at the present stage of development of QKD. 

Among the assumptions made in most unconditional security proofs, one is manifestly at odds with the behavior of a real device: namely, the fact that bounds are usually provided only in the asymptotic limit of infinitely long keys. On this issue, no convergence is possible unless the theorists make the effort of developing \textit{finite-key analysis}. Remarkably, all the elements for a rigorous finite-key analysis were already present in the very first unconditional security proof by Mayers \cite{may96}. However, his work was too innovative and also too complex to be duly appreciated. His subsequent work with Inamori and L\"utkenhaus \cite{ina07} went also rather unnoticed; moreover, it was shown later that their approach does not yield composable security \cite{BHLMO05,KRBM07} and must therefore be abandoned. Other partial estimates showed that the finite-key correction is quite important in the usual range of operation of QKD systems \cite{LoChauArdehali,Ma05,Wang05,mey06}.

The first study, in which finite-key analysis is integrated in a proof of composable unconditional security, is Hayashi's analysis of the BB84 protocol with decoy states~\cite{hay2}. This is, to our knowledge, the only finite-key bound to have been applied to experimental data as of today \cite{hase}. Independently, Renner and one of us also developed security proofs in the non-asymptotic limit \cite{scaren1,scaren2} based on the formalism developed in Ref.~\cite{rennerthesis}. In the present paper, we use this approach to derive explicit finite-key security bounds for practical implementations of the BB84 coding. In Section \ref{sec2}, we provide the general elements of finite-key formalism following Refs \cite{scaren1,scaren2}. In Section \ref{sec3} we apply these tools to one-way prepare-and-measure implementations of BB84 with weak coherent pulses, both without and with decoy states: we derive an unconditional security bound for the first and a partial bound for the second. Part of the results overlap with those of Hayashi and co-workers \cite{hay3}. In Section \ref{sec4} we repeat the same study for entanglement-based implementations of the BB84 coding, i.e. for the Bennett-Brassard-Mermin 1992 (BBM92) protocol \cite{bbm92}.

\section{Finite-key formalism}
\label{sec2}

\subsection{Asymmetric BB84 protocol}

We consider the BB84 coding with asymmetric role of the bases \cite{LoChauArdehali}: the key is obtained from the events in which both Alice and Bob have used the $Z$ basis, while the correlations in the $X$ basis are used to estimate Eve's knowledge. We write $p_Z$ the probability that the $Z$ basis is chosen and $p_X=1-p_Z$ the probability that the $X$ basis is chosen (to keep things simple in this general survey, we assume that these probabilities are the same for Alice and Bob). Therefore, denoting $N$ the length of Alice's and Bob's lists before sifting (basically, the number of signals detected by Bob), the raw key will be of length $n=Np_Z^2$, Eve's information is estimated on a sample consisting of $m=Np_X^2$, and $2Np_Zp_X$ signals are discarded in sifting. We denote by $\mathbf{e_Z}$ and $\mathbf{e_X}$ the measured error rates in the two bases (in the whole paper, we use boldface fonts for the quantities that are directly measured in the protocol).

\subsection{Finite-key bound for the secret fraction}
\label{fkgeneral}

Although the finite-key formalism has been generalized to accommodate more general forms of classical post-processing \cite{scaren2}, in this paper we consider the extraction of a secret key through one-way post-processing without pre-processing. Out of the $n$ pairs of bits that form the raw key, Alice and Bob want to extract a secret key of length $\ell\leq n$. We refer to the ratio $r=\ell/N$ as to the \textit{secret fraction}. The asymptotic value of $r$ is given by the well-known Devetak-Winter bound \cite{DevWin05}
\ba \lim_{N\rightarrow\infty}r &=& S(X|E) - H(X|Y) \, \label{devwin} \ea
where $S(X|E) := S(X E) - S(E)$ and $H(X|Y) := H(X Y) - H(Y)$ are the conditional von Neumann and Shannon entropies, respectively, evaluated for the joint state of Alice and Bob's raw key and the system controlled by Eve (after the sifting step). The main result of Refs \cite{scaren1,scaren2} says that the finite key version of this bound can also be cast in a rather simple form, namely
\ba r &=&p_Z^2\,\left[S_{\xi}(X|E)-\Delta(n)-\leak_{\EC}\right]\label{raten}\ea
whose terms we are going to comment:
\begin{itemize}
\item The first correction to the asymptotic bound is the factor $n/N=p_Z^2$. Its meaning is pretty obvious: only $n$ signals out of $N$ form the raw key. In the limit $N\rightarrow\infty$, one can choose $p_Z\rightarrow 1$ because a small fraction of signals will give an accurate enough estimation of the parameters --- typically, $m\propto \sqrt{N}$ i.e. $p_X\propto N^{1/4}$ \cite{scaren1,hay4}; see also our study below.

\item The second correction is the one represented by the notation $S_{\xi}(X|E)$, the modification of Eve's uncertainty on single copies $S(X|E)$. Its meaning is also obvious. Eve's information is estimated using measured parameters, e.g. error rates. In a finite key scenario, these parameters are estimated on samples of finite length: therefore, one has to allow for statistical fluctuations.

Specifically, let $\lambda$ be one of the parameters that enter Eve's information (to fix ideas, think to $e_X$); and let $d$ be the number of outcomes of a POVM needed to estimate it (for error rates of bits, $d=2$ since the outcomes are ``Alice$=$Bob'' and ``Alice$\neq$Bob''). Suppose then that $m'$ signals have been used to estimate $\lambda$: then the deviation of the estimate $\lambda_{m'}$ from the ideal estimate $\lambda_{\infty}$ can be quantified by
\ba
|\lambda_{m'}-\lambda_{\infty}|&\leq& \xi(m',d)\,=\, \sqrt{\frac{\ln(1/\eps_{\PE})+d\ln(m'+1)}{2m'}}\label{eqpe}
\ea
where $\eps_{\PE}$ is the failure probability of the parameter estimation\footnote{The law of large numbers we are using reads $\sum_{k=1}^d|\lambda_m(k)-\lambda_{\infty}(k)|\equiv \sum_{k=1}^d|\Delta_m(k)|\leq \sqrt{[2\ln(1/\eps_{\PE})+d\ln(m+1)]/m}$ \cite{cover}. The constraint $\sum_{k=1}^d\lambda_m(k)=\sum_{k=1}^d\lambda_{\infty}(k)=1$, i.e. $\sum_{k=1}^d\Delta_m(k)=0$, implies that the deviation for the parameter $\lambda=\lambda(1)$ that we want to estimate is given by Eq.~(\ref{eqpe}) --- more precisely, Eq.~(\ref{eqpe}) is exact for $d=2$, while for $d>2$ it represents the largest possible deviation. The factor $\demi$ was missing in previous works \cite{scaren1,scaren2}, therefore the lower bounds presented there may be made slightly more optimistic. After inspection, the net result is that the curves obtained for $N$ can actually be obtained already for $N'\sim N/2$.}. We shall write the upper and the lower bounds compatible with the fluctuations as
\ba
\lambda^U\,=\,\min(\lambda+\xi,1)&,&\lambda^L\,=\,\max(\lambda-\xi,0)
\ea
because all the $\lambda$'s estimated below are probabilities (error rates, fraction of multi-photon pulses etc). In all that follows, for simplicity of notation we shall omit the $\max$ and $\min$.

We stress that the notation $\lambda^{U,L}$ was first introduced in \cite{Ma05}. Here the expressions are different, since they considered relative errors drawn from a normal distribution, while our estimate (\ref{eqpe}) quantifies absolute errors and does not assume any specific form for the underlying distribution. This is a requirement of the finite-key formalism we are using. This difference will lead to some minor discrepancies with previously published works, see Section \ref{secdecoy}. The possibility of rephrasing the formalism in terms of relative errors is listed among the open issues at the end of this paper.

\item The third correction to be commented is
\ba
\Delta(n)&=&7\sqrt{\frac{\log_2(2/\bar{\varepsilon})}{n}}\,+\,\frac{2}{n}\log_2(1/\varepsilon_{\PA})\,.\label{Delta}
\ea This numerical term is all that is left of the technicalities of unconditional security proofs. We give here only a very rapid sketch of its origin and refer to \cite{scaren1,scaren2} for all details. Eve's uncertainty is quantified by a generalized conditional entropy called smooth min-entropy and denoted $H^{\bar{\varepsilon}}_{min}(X^{(n)}|E^{(N)})$. The parameter $\bar{\varepsilon}$ quantifies the ``smoothing'': it is a parameter of the theory, whose value can be optimized numerically (see below).

The smooth min-entropy cannot be computed because it is virtually impossible to parametrize the most general state $\rho_{X^nY^n E^{(N)}}$ compatible with the few observed parameters. In a first step therefore, one estimates the deviation that is obtained assuming that the state consists of $n$ independent realizations of a given single-copy state, i.e. $\rho_{X^nY^n E^{(N)}}=(\sigma_{XYE})^{\otimes n}$. In general, this estimate requires a de Finetti-type theorem \cite{rennerthesis}, which leads however a very pessimistic overhead in finite-key analysis (though a recent new approach should provide a much tighter estimate \cite{chri08}). For BB84 however, it turns out that no deviation is expected at all: because of the symmetry of the protocol, the state can be written as a convex combination of products of Bell states without loss of generality \cite{KGR,gotlo}. The product form of the state being thus justified, it can further be proved that the smooth min-entropy is lower bounded by $n[S_{\xi}(X|E)-\delta]$, where $\delta$ is the first term of the sum in (\ref{Delta}). The second term in the sum comes from the fact that, in the non-asymptotic case, the task of privacy amplification itself may fail with probability $\varepsilon_{\PA}$.

\item Finally, $\leak_{\EC}$ replaces $H(X|Y)$ as the fraction to be removed in error correction. It is also well-known that practical error correction codes do not reach the Shannon limit. Typically,
\ba
\leak_{\EC}&\approx&f_{\EC}H(X|Y)+\frac{1}{n}\log_2(2/\eps_{\EC})
\ea
where $f_{\EC}>1$ depends on the code and $\eps_{\EC}$ is the failure probability of the error correction procedure. In a practical implementation, this quantity is a direct outcome of running the error-correcting code (although one must be careful in case a two-way error correction code is actually used \cite{lut99}). 

\end{itemize}

Even if everything has been carried out ``perfectly'', there is no such thing as perfect security. In our formalism, the security parameter $\eps$ has an operational meaning: it represents the maximum probability failure that is tolerated on the key extraction protocol (for instance, $\eps=10^{-10}$ can be loosely read as: ``one can distribute $10^{10}$ keys before something may go wrong''). With this interpretation, it is clear that the total security parameter is simply the sum of the probabilities of failures of each procedure described above, so that \ba\eps&=&\eps_{\EC}+\bar{\eps}+n_{\PE}\,\eps_{\PE}+\eps_{\PA}\label{epssum}\ea where $n_{\PE}$ is the number of parameters that must be estimated (for simplicity, we set all the corresponding $\eps_{\PE}$ as equal).

\subsection{Putting finite-key bounds into practice}
\label{sec2c}

In the previous paragraph, we have sketched the elements that enter the calculation of the secret fraction $r$ for BB84 coding in a finite-key scenario. A few remarks are needed to complete the picture. First of all, the performance of an implementation is not quantified by $r$ alone, but by the \textit{secret-key rate}
\ba
K&=&\mathbf{R}\,r
\ea
where $\mathbf{R}$ is the detection rate. In this paper, we use rates per sent qubit; the usual rates per second are obtained by multiplying our results with the frequency at which the source is operated.

An actual experiment is described by the following parameters:
\begin{itemize}
\item The user must set his/her desired bound $\eps$ on the total failure probability of the key distribution task: how often is one willing to tolerate that the final outcome of the post-processing is \textit{not} a perfect secret key.
\item The post-processing code determines the size of the blocks on which privacy amplification is applied. This is the exact meaning of the parameter $n$: the length of the raw key as it is processed. Indeed, the raw key itself can be made longer by running the experiment for a longer time, but this mere fact cannot increase the security if the data are sliced and processed in blocks.
\item The choice of an error correcting code determines $\leak_{\EC}$, i.e. $f_{\EC}$ and $\eps_{\EC}$. 
\end{itemize}

All the other parameters can be chosen to optimize $K$. The three auxiliary security parameters $\bar{\eps}$, $\eps_{\PE}$ and $\eps_{\PA}$ are necessary in the derivation of the bound but need not be specified by the user. Their value can be optimized at the moment of computing $r$, under the constraints of being positive and satisfying (\ref{epssum}). The parameters that enter in the design of the experiment, however, must obviously be chosen before the experiment is run. Explicitly, the flow of operations goes as follows:
\begin{enumerate}

\item Find $n$, $f_{\EC}$ and $\eps_{\EC}$ as given by the chosen post-processing code; choose $\eps$.

\item Provide \textit{a priori} expected values of the parameters that are going to be measured: detection rate $R$, error rate in either basis $e_X$ and $e_Z$, and others. Insert these expressions in the finite-key bound and optimize the design of the experiment: i.e. find the values of the light intensity $I$, of $p_X$ and possibly of other quantities, that maximize $K$.

\item Run the experiment.

\item Insert the \textit{measured} values $\{\mathbf{R},\mathbf{e_X},\mathbf{e_Z},...\}$ in the finite-key bound and run again the optimization of $r$ over the $\eps$'s but using the value of $I$, $p_X$ etc. used in the experiment --- which may not be optimal for the measured values, especially if these differ significantly from the expected ones. This gives how much privacy amplification must be performed.

\item Run classical post-processing and obtain the secret key.

\end{enumerate}

The procedure we have just sketched has been implicitly assumed in many previous papers, but to our knowledge has not been explicitly spelled out before. It is therefore worth while elaborating more on it, at the risk of some redundancy. Consider for instance the intensity $I$ of the light source: it must obviously be chosen before the experiment is run. This choice involves an optimization between two effects: on the one hand, the detection rate (so the raw key length) will increase linearly with $I$; on the other hand, high $I$ lead to some nuisances (e.g. Eve's information increases in prepare-and-measure schemes, or the error rate increases in entanglement-based schemes; see later). In order to find the optimal value of $I$, one has to provide some \textit{a priori} expected expressions of the detection rate, Eve's information, error rate... as functions of $I$. For instance, if, at the calibration stage, the transmission of the quantum channel and the efficiency of the detectors have been measured to be, respectively, $t$ and $\eta$; then \textit{a priori} one expects $R\approx I\,t\eta$.

Now, once the experiment is run, there is no guarantee that the measured $\textbf{R}$ will be equal, or even close, to $R$: Eve's attack may introduce many more losses than expected. Actually, anything can happen: for instance, in an entanglement-based scheme, one may observe that the error rate does not vary with the intensity, if Eve decides to block all the multiple-pair pulses. We don't know \textit{why} Eve would do that, just as we do not question why she has introduced a given amount of error and not more or less: the only thing we must ensure is that, given the \textit{measured} parameters, Eve's information is always upper-bounded. Of course, the value of $I$ that we have chosen, and that would have been optimal in the expected condition, may turn out to be seriously sub-optimal given the measured values. But again, this is perfectly fine: it just means that Eve's attack is too strong for any secrecy to be extractable.

In this paper, we take care of distinguishing clearly the security bounds, always formulated in terms of measured quantities and therefore applicable to any experiment, from the derived numerical bounds obtained using some \textit{a priori} expected values.

In what follows, we provide the finite-key bounds (both the general expression and its numerical evaluations for \textit{a priori} expected values) for different practical implementations of the BB84 coding.

\section{Prepare-and-measure implementations with weak coherent pulses}
\label{sec3}

\subsection{Asymptotic bounds}

\subsubsection{Generalities}

We consider a source producing a train of weak coherent pulses of average intensity $\mu$; the following analysis is valid provided no phase coherence between successive pulses \cite{lopre}. In this case, the signal sent by Alice can equivalently be described as a Poissonian distribution of Fock states, such that the probability of sending a $k$-photon pulse is
\ba
p_A(k|\mu)&=&e^{-\mu}\,\frac{\mu^k}{k!}\,.
\ea
Asymptotic bounds for unconditional security of such implementations have been derived using several approaches \cite{gllp,fun06a,kra07}; we refer to these papers and to Section IV of Ref.~\cite{sca08} for all details. Without loss of generality, one can assume that (i) Eve learns the number of photons in each pulse and adapts her strategy to it, and (ii) Eve forwards single-photon signals to Bob. An important step in such proofs is the reduction, or ``squashing'', of the state of the physical signal into a qubit. Specifically, one assumes that the measurement performed by the photon counters can be described by first squashing the signal on a finite-dimensional Hilbert space, then performing a measurement in this space \cite{gllp}. When those proofs were proposed, the squashing property of detectors was conjectured; recently, this property has been proved to hold in the case of BB84 \cite{bea08,tsu08}.

The probability that Bob detects something, given that the pulse contained $k$ photons, is given by $p_B(k|\mu)\,=\,p_A(k|\mu)\,f_k$, where $f_k$ is the probability that Eve forwards a photon to Bob. Note that all the losses, both those due to the transmission line and those due to the detector efficiency, are included in $f_k$ and are therefore given to Eve: this is the so-called uncalibrated-device scenario, the only one in which unconditional security can be proved as of today \cite{sca08,bea08} and also justified by some clever realistic attacks \cite{mak08}. The $p_B(k|\mu)$ are submitted to the constraint that their sum must match the total observed detection rate: \ba\mathbf{R}&=&\sum_k p_A(k|\mu)\,f_k\,.\label{constr1}\ea It is customary to write
\ba
Y_k(\mu)&=&\frac{p_A(k|\mu)\,f_k}{\mathbf{R}}\,.
\ea
Also, on $k$-photon pulses, Eve introduces the error rate $e_{X,Z}(k)$ in either basis. The measured error rates constrain these parameters to satisfy
\ba
\mathbf{e_{X,Z}}&=&\sum_k Y_k(\mu)\,e_{X,Z}(k)\,.\label{constr2}
\ea
The set of $f_k$ and $e_{X,Z}(k)$ fully parametrize Eve's attack.

Finally, under the additional assumption that Alice's and Bob's raw keys have maximal entropy (i.e. that the bit values 0 and 1 both occur with probability 1/2), the asymptotic expression for $S(A|E)$ for a given choice of $\mu$ is
\ba
S(A|E)&=&\min_{\mathrm{Eve}}\,\left\{Y_0(\mu)\,+\,Y_1(\mu)\left[1-h\big(e_X(1)\big)\right]\right\}\label{saegen}
\ea
where $h$ is binary entropy and the minimum must be taken over all possible choices of the $f_k$ and the $e_{X,Z}(k)$ compatible with the measured parameters. Note that $\mathbf{e_Z}$ does not appear in Eve's information: this is a consequence of the fact that Eve's information on the $Z$ basis is a function of the error introduced in the complementary basis\footnote{As well-known, one must be careful in using this intuitive argument: in the case of the six-state protocol, for instance, $\mathbf{e_Z}$ does enter in the expression of Eve's information even for an asymmetric implementation, see e.g. Appendix A of Ref.~\cite{sca08}.}. Therefore, in discussing $S(A|E)$ and its finite key correspondent $S_{\xi}(A|E)$, we don't mention $\mathbf{e_Z}$ any more.

\subsubsection{Implementations without decoy states}

In the case of implementations \textit{without decoy states}, the optimal choice of parameters is given by $f_0=0$, $f_k=1$ and $e_{X}(k)=0$ for $k\geq 2$; the estimates $\tilde{Y}_1(\mu)$ and $\tilde{e}_X(1)$ are therefore fully determined by (\ref{constr1}) and (\ref{constr2}), leading to
\ba
S(A|E)\,=\,\tilde{Y}_1(\mu)\left[1-h\big(\tilde{e}_X(1)\big)\right]\,,&\mathrm{with}& \tilde{Y}_1(\mu)=1-\frac{p_A(k\geq 2|\mu)}{\mathbf{R}}\quad \mathrm{and}\quad \tilde{e}_X(1)=\frac{\mathbf{e_X}}{\tilde{Y}_1(\mu)}\label{saewithout}
\ea
where obviously $p_A(k\geq 2|\mu)=1-e^{-\mu}(1+\mu)$\,.

\subsubsection{Implementations with decoy states}

Implementations \textit{with decoy states} aim at estimating the $f_k$ and $e_{X}(k)$ more directly \cite{hwa03,wan05,lo05}. For each pulse, Alice picks at random an intensity $\mu\in\{\mu_{\gamma}\}_{\gamma\in\Gamma}$ from a set of possible values (the protocol should specify \textit{which} are these values and with which probability $q_\gamma$ each one is chosen, but of course not which one will be used for each pulse). For the items in which Bob announces a detection, Alice reveals which $\mu_{\gamma}$ was used; she and Bob can therefore estimate parameters conditioned on this information. However, the parameters $f_k$ and the $e_{X,Z}(k)$ that define Eve's attack must be the same for all $\mu_{\gamma}$. Therefore, the constraints (\ref{constr1}) and (\ref{constr2}) become a set of $2|\Gamma|$ constraints
\ba\mathbf{R^{\gamma}}&=&\sum_k p_A(k|\mu_\gamma)\,f_k\,,\label{constr1decoy}\\ \mathbf{e^{\,\gamma}_{X}}&=&\sum_k Y_k(\gamma)\,e_{X}(k)\label{constr2decoy}\ea where $Y_k(\gamma)=p_A(k|\mu_\gamma)\,f_k/\mathbf{R^{\gamma}}$. Through this method, Eve's attack can in principle be exactly parametrized \cite{lo05}, but this requires $|\Gamma|=\infty$. However, only $f_0$, $f_1$ and $e_X(1)$ enter the expression (\ref{saegen}) of $S(A|E)$, and it is evident that a pretty good estimate is already obtained with a few values of $\mu_{\gamma}$ \cite{wan05}. Asymptotically, 
\ba
S(A|E)\,\equiv \,S(A|E,\bar{\gamma})&=&\tilde{Y}_0({\bar{\gamma}})\,+\, \tilde{Y}({\bar{\gamma}})\,\left[1-h\big(\tilde{e}_X(1)\big)\right]\label{saewith}
\ea
where $\tilde{Y}_k(\gamma)=p_A(k|\mu_\gamma)\,\tilde{f}_k/\mathbf{R^{\gamma}}$ and where $\bar{\gamma}$ is defined as the value of $\gamma$ that maximizes $K_\gamma=\mathbf{R^{\gamma}}\left[S(A|E,\gamma)-h(\mathbf{e^{\gamma}_{Z}}) \right]$. This is the case because, in the asymptotic regime, one can set $q_{\bar{\gamma}}\rightarrow 1$ and use the other intensities in a negligible fraction of cases. In the finite-key regime, this can no longer be the case: below, for simplicity, we shall consider the case where the key is extracted only out of one of the intensities.

\subsubsection{An example of decoy states}
\label{ssex3}

For the explicit finite-key study below, we consider a specific choice of decoy state implementation, first studied in the pioneering paper by Wang \cite{wan05}. The protocol uses \textit{three intensities}, one of which is actually $\mu_\emptyset=0$, while the other two are denoted $\mu_{\mathrm{I}}$ and $\mu_{\mathrm{II}}$ (we note here that, in theory, the condition $\mu=0$ seems trivial to realize: just shut down the power or put an obstacle in the light path; but if the pulsing rate is required to be high, i.e. if the switch has to operate with high speed, it may be actually very difficult to shut down the power completely). The relations $\mu_{\mathrm{I}}\leq \mu_{\mathrm{II}}$ and $\mu_{\mathrm{I}}e^{-\mu_{\mathrm{I}}}\leq \mu_{\mathrm{II}}e^{-\mu_{\mathrm{II}}}$, i.e. $p_A(0|\mathrm{I})\geq p_A(0|\mathrm{II})$ and $p_A(1|\mathrm{I})\leq p_A(1|\mathrm{II})$, are assumed to be valid.

When $\mu=\mu_{\emptyset}$, all the pulses are empty so $p_A(k|\emptyset)=\delta_{k,0}$ and one immediately obtains the estimates
\ba
\tilde{f}_0\,=\,\mathbf{R^{\emptyset}}&,& \tilde{e}_{X}(0)\,=\,\mathbf{e^{\,\emptyset}_{X}}\,.
\ea The estimate for $f_1$ can be extracted using either $\mathbf{R^\gamma}=p_A(0|\mu_\gamma)\,\tilde{f}_0+p_A(1|\mu_\gamma)\,\tilde{f}_1+\mathbf{R^\gamma}\Delta^\gamma $ where $\Delta^\mathrm{I}$ and $\Delta^\mathrm{II}$ are given respectively by Eqs. (13) and (15) of \cite{wan05}; explicitly
\ba
\tilde{f}_1&=&\frac{1}{\mu_{\mathrm{II}}-\mu_{\mathrm{I}}}\left[ \mathbf{R^{\mathrm{I}}} \frac{\mu_{\mathrm{II}}}{p_A(1|{\mathrm{I}})} - \mathbf{R^{\mathrm{II}}} \frac{\mu_{\mathrm{I}}}{p_A(1|{\mathrm{II}})}\right]\,-\, \tilde{f}_0\,\frac{\mu_{\mathrm{II}}+\mu_{\mathrm{I}}}{\mu_{\mathrm{II}}\mu_{\mathrm{I}}}\,.\label{f1decoy}
\ea
To obtain an estimate for $e_X(1)$, we note that (\ref{constr2decoy}) becomes $\mathbf{e^{\,\gamma}_{X}}=\tilde{Y}_0(\gamma)\,\tilde{e}_{X}(0)+\tilde{Y}_1(\gamma)\,\tilde{e}_{X}(1)+Y_\Delta(\gamma)\,\tilde{e}_{X}(\Delta,\gamma)$ where $Y_\Delta(\gamma)=\Delta^\gamma/\mathbf{R^\gamma}$. Now, the two $\tilde{e}_{X}(\Delta,\gamma)$ depend on $\gamma$ and are unknown, but must be non-negative; this implies that the largest value of $\tilde{e}_{X}(1)$ is
\ba
\tilde{e}_{X}(1)&=&\min_{\gamma\in\{\mathrm{I},\mathrm{II}\}}\,\left(\frac{\mathbf{e^{\,\gamma}_{X}}-\tilde{Y}_0(\gamma)\,\tilde{e}_{X}(0)}{\tilde{Y}_1(\gamma)}\right)\,.
\ea

\subsection{Finite-key security bounds}

In the previous paragraph, we have collected the necessary notations and the known asymptotic bounds. Note that the only quantity that varies according to the implementation is $S_{\xi}(X|E)$ and the recipe to obtain it from the known asymptotic bounds $S(X|E)$ is straightforward: replace the estimate of each parameter by its worst-case value compatible with the deviation $\xi(m',d)$ given in (\ref{eqpe}). Here we derive $S_\xi(A|E)$ from $S(A|E)$, both for implementations without and with decoy states.

\subsubsection{Implementations without decoy states: unconditional security bound}

We have to identify which parameters are subject to statistical fluctuations among those that enter in Eq.~(\ref{saewithout}): \begin{itemize} \item First we notice that $\mathbf{R}$ is just the number of signals detected by Bob $N$ divided by the number of signals sent by Alice, in the given run of the experiment. No statistical estimate is involved, therefore there is no fluctuation here. This statement may seem surprising. To understand it fully, one must come back to the difference between measured values and \textit{a priori} expected values (end of Section \ref{sec2c}). Indeed, the \textit{expected value} $R\approx \mu t\eta$ will surely be subject to fluctuations; but this just means that the observed value of $\mathbf{R}$ may differ from $\mu t\eta$. When assessing security, however, one must plug the \textit{measured} value, and there is no reason to burden this value with a fluctuation.

\item The fraction $\tilde{Y}_1(\mu)$ is an estimate of the fraction of signals that reach Bob arising from a single-photon pulse; it depends explicitly on the probability that Alice's pulse contains more than two photons, and this quantity is obviously subject to fluctuations (by ``bad luck'', Alice might have sent out only two-photon pulses!). All the $N$ signals are involved in this estimate, which could in principle be done with a 2-outcomes POVM (``$k<2$'' versus ``$k\geq 2$''). Therefore, with probability $1-\eps_{\PE}$, the real $p_A(k\geq 2)$ differs from the expected one $p_A(k\geq 2|\mu)$ at most by $\xi(N,2)$.

\item The real error rate in $X$ basis may deviate from the observed fraction of wrong events $\mathbf{e_X}$; because $m$ signals are used for the measurement, the deviation is bounded by $\xi(m,2)$. 

\end{itemize}

In summary, there are two parameters subject to fluctuations ($n_{\PE}=2$) and \ba
S_{\xi}(A|E)&=&Y_1^{L}(\mu)\left[1-h\big(e^{U}_X(1)\big)\right]\label{saexiwithout}\ea with
\ba
Y_1^{L}(\mu)\,=\,1-\frac{[1-e^{-\mu}(1+\mu)]+\xi(N,2)}{\mathbf{R}}& \mathrm{and} & e^{U}_X(1)\,=\, \frac{\mathbf{e_X}+\xi(m,2)}{Y^{L}_1(\mu)}\,.
\ea
Note that $Y_1^{L}(\mu)=\tilde{Y}_1(\mu)-\frac{\xi(N,2)}{\mathbf{R}}$ and $e^{U}_X(1)\approx \frac{\mathbf{e_X}+\xi(m,2)}{\tilde{Y}_1(\mu)}+\frac{\xi(N,2)}{\mathbf{R}}=\tilde{e}_X(1)+\frac{\xi(m,2)}{\tilde{Y}_1(\mu)}+\frac{\xi(N,2)}{\mathbf{R}}$. In particular, \textit{two} finite-statistics effects provide corrections to the estimate of $e_X(1)$: the fact that the total error rate $\mathbf{e_X}$ was estimated on $m$ samples and the fact that the fraction of single-photon pulses was inferred from $N$ samples.

\subsubsection{Implementations with decoy states: approximate bound}

For decoy states protocols, three parameters have to be estimated, namely $f_0$, $f_1$ and $e_X(1)$; so $n_{\PE}=3$. The recipe to obtain $S_\xi(A|E)$ from $S(A|E)$ is:
\begin{itemize}
\item In the first constraint (\ref{constr1decoy}), one introduces fluctuations to the $p_A(k|\mu_\gamma)$, then solves the system of equations for the measured values $\mathbf{R^{\gamma}}$ and obtains the finite-key estimates for the $f_k$;
\item One inserts these estimates into the second constraint (\ref{constr2decoy}), adds the fluctuations to the estimated error rates $\mathbf{e^{\,\gamma}_{X}}$ and solves for the $e^U_X(k)$.
\end{itemize}
While this second step is easy to implement, the first one is much harder and its full treatment goes beyond the scope of this paper\footnote{Let us mention one of the reasons for such a complexity: while one has to consider $f_0^L$ and $f_1^L$ because of (\ref{saewith}), it is not evident which fluctuation should be retained for the $f_{k\geq 2}$. In other words, given that the eavesdropper is allowed to take advantage of deviations from the Poissonian behavior, it is hard to quantify how Eve is going to redistribute the fluctuations removed from $f_0$ and $f_1$ among the other $f_k$'s.}. Here we follow a simpler recipe: we solve first (\ref{constr1decoy}) without fluctuations, obtain the expressions for $f_0$ and $f_1$, then add a fluctuation to the $Y_{k}(\gamma)=p_A(k|\gamma)f_k$. Of course, having opted for this simplified treatment, we cannot claim unconditional security for the derived bound.

We particularize directly to the three-intensity protocol sketched above (\ref{ssex3}). Since the zero-pulse fractions $Y_0(\gamma)$ are estimated using only $\mu_\emptyset=0$, and the POVM can be rendered by the two outcomes ``detection'' versus ``no-detection'', we have
\ba
Y_0^L(\gamma)&=&\left[p_A(0|\gamma)\mathbf{R^\emptyset}-\xi(N_\emptyset,2)\right]/\,\mathbf{R^\gamma}\,.\label{y0}\ea Similarly, once the parameter $f_1$ is estimated as (\ref{f1decoy}), we obtain
\ba
Y_1^L(\gamma)&=&\left[p_A(1|\gamma)\,\tilde{f}_1-\xi(N_\gamma,2)\right]/\,\mathbf{R^\gamma}\label{y1}\ea because all the $N_\gamma$ signals are involved in the virtual two-outcome POVM ``less than two photons'' versus ``two and more photons''. Finally, the recipe to obtain $e^{U}_{X}(1)$ is the usual one: insert the finite estimates $Y^L_k(\gamma)$ and increase the measured error rates by the corresponding fluctuations. For this last term, however, two points are worth noting. First, the worst case fluctuation is the one that reduces $\mathbf{e^{\,\emptyset}_{X}}$, because this amounts at increasing $e^{U}_{X}(1)$. Second, all the $N_\emptyset$ events can be used to estimate this error rate: obviously, if Alice's pulse is empty, there is no difference between encoding in $X$ or in $Z$; so Bob can assume that he has always used the ``right'' basis to measure these signals. All in all,
\ba
e^{U}_{X}(1)&=&\min_{\gamma\in\{\mathrm{I},\mathrm{II}\}}\,\left(\frac{e^{\gamma,U}_{X}- Y^L_0(\gamma)\,e^{\emptyset,L}_{X}}{Y^{L}_1(\gamma)}\right)\,.\label{ex1}
\ea
with $e^{\gamma,U}_{X}=\mathbf{e^{\,\gamma}_{X}}+\xi(m_\gamma,2)$ and $e^{\emptyset,L}_{X}=\mathbf{e^{\,\emptyset}_{X}}-\xi(N_\emptyset,2)$.

\subsection{\textit{A priori} expected values for experiment design}

For simplicity in this paper we plot curves for a fixed value of $N$, the length of the unsifted key\footnote{We mentioned in \ref{sec2c} that the parameter that really define an experiment is $n$ (the size of the blocks on which post-processing is applied) and not $N$. Of course, one could in principle run optimizations for fixed $n$; but this requires the introduction of additional assumptions. For instance, if only $n$ is fixed and one sets $N=n/p_Z^2$, then the obvious optimal is $p_Z=0$ i.e. $N=\infty$ signals are used, most of them to estimate the parameters. To avoid such situations, one may set $p_X\leq p_Z$. However, leaving aside that this choice is \textit{a priori} arbitrary, the situation becomes even more complicated in decoy states: for instance, one must make sure that none of the intensities is used infinitely many times. To avoid such complications, we find it more clear in this paper to keep the number of detected quantum signals fixed. A posteriori, one always find $n=Np_Z^2\approx N-O(\sqrt{N})$.}. The expected values that we choose for our \textit{a priori} expected values depend on the parameters $t$, the transmittivity of the channel Alice-Bob, $\eta$ and $p_d$, the quantum efficiency and the dark count rate of Bob's detectors respectively.

The expected value of the detection rate we use is\footnote{In the expression of $R(\mu)$, we have neglected the contribution of double-clicks. This does not mean that double-clicks can just be neglected in an implementation (more in Section \ref{sec4}). Actually, since our bounds are based on squashing, they must be replaced by a random bit and therefore contribute in a similar way as the dark counts. We neglect in the \textit{a priori} expected values because their contribution is numerically small.}
\ba
R(\mu)&=&1-(1-2p_d)\,e^{-\mu t\eta}
\ea
Accordingly, error rates will be assumed to take the form
\ba
e_Z(\mu)\,=\,e_X(\mu)&=& \frac{\left(1-e^{-\mu t\eta}\right)Q\,+\,e^{-\mu t\eta}p_d}{R(\mu)}
\ea 
where $Q$, often called optical quantum bit error rate, is the error induced by the channel; in a depolarizing channel with visibility $V$, the BB84 coding leads to $Q=(1-V)/2$.

\subsubsection{Implementations without decoy states}

\begin{figure}[ht]
\includegraphics[scale=0.6]{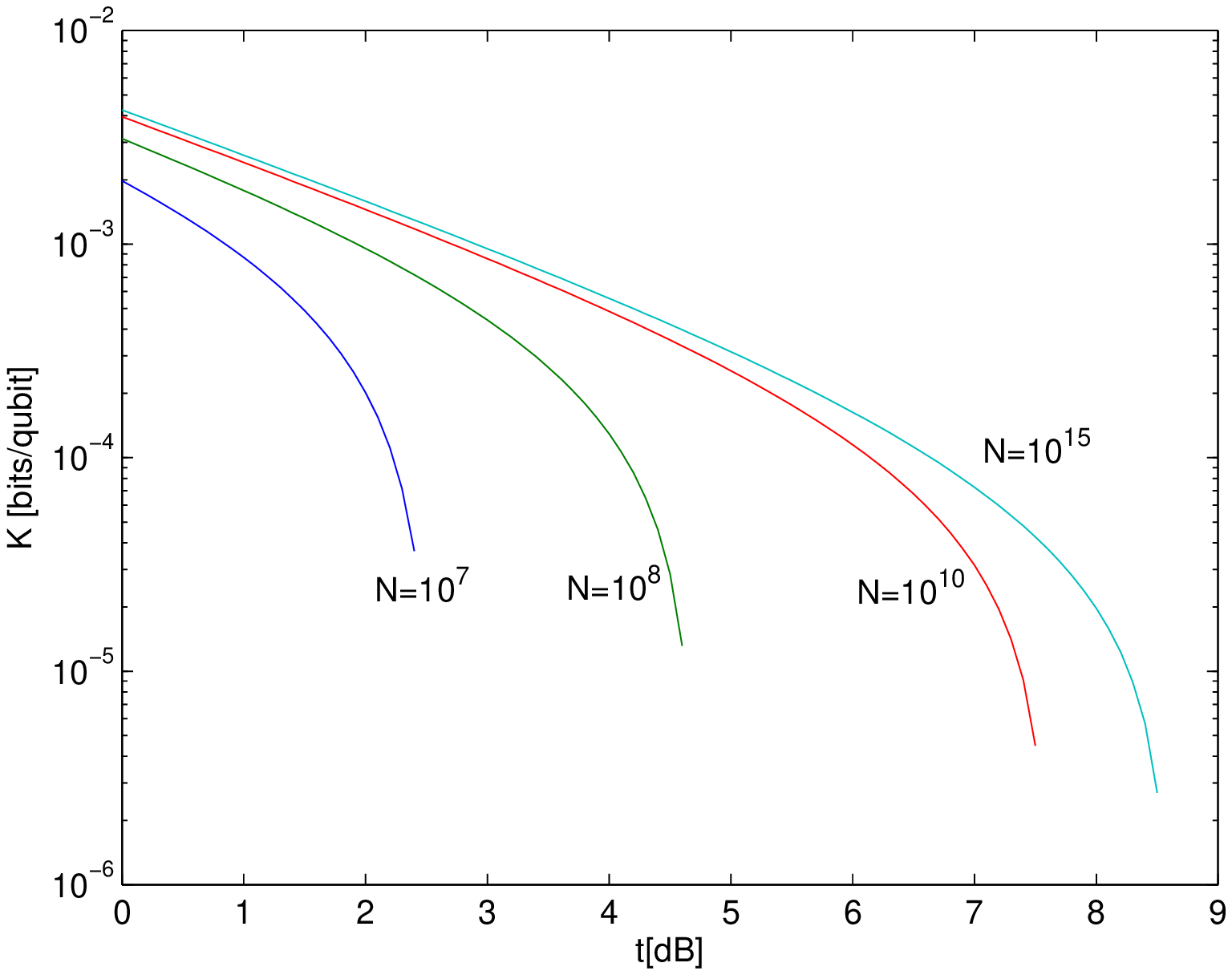}\\
\includegraphics[scale=0.6]{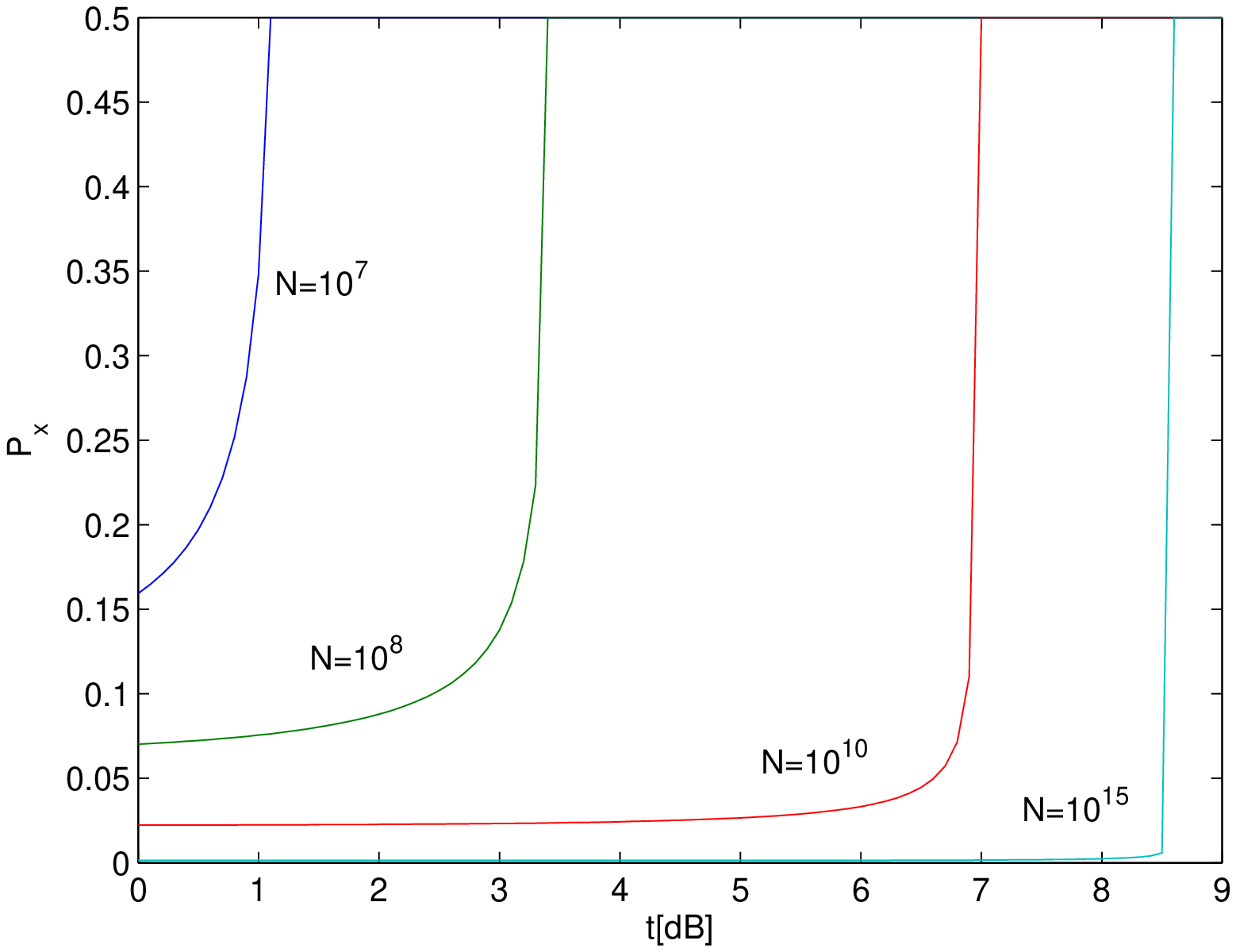}
\caption{Finite-key study of implementations of BB84 with weak coherent pulses, without decoy states. As a function of the transmittivity of the channel $t$: upper graph, secret key rate $K$ from eq.~(\ref{boundwithout}); lower graph: corresponding optimal value of $p_X$. Parameters: $\eps=10^{-5}$, $\eps_{\EC}=10^{-10}$, $\mathrm{leak}_{\EC}(e)=1.05\,h(e)$, $Q=0.5\%$, $\eta=0.1$, $p_d=10^{-5}$.}\label{figwithout}
\end{figure}

We consider first implementations without decoy states. We have to optimize 
\ba
K&=&R(\mu)\,p_Z^2\,\left[S_\xi(A|E)-\Delta(n)-\mathrm{leak}_{\EC}(e_Z)\right] \label{boundwithout}
\ea
for $S_\xi(A|E)$ given in (\ref{saexiwithout}), over $\mu$ and over the finite-key parameters. The result is shown in Fig.~\ref{figwithout} for a choice of parameters corresponding to today's state-of-the-art. We see that at least $N\approx 10^7$ signals are required to extract a secret key. As for the optimal parameters: $\mu$ is found to be very close to the well-known value $t\eta$ \cite{lut99,sca08} irrespective of $N$; far from the critical distance, $p_X$ is constant with the transmittivity and varies as $N^{-1/4}$, whence $m\sim \sqrt{N}$.

\subsubsection{Implementations with decoy states: case study}
\label{secdecoy}

\begin{figure}[ht]
\includegraphics[scale=0.6]{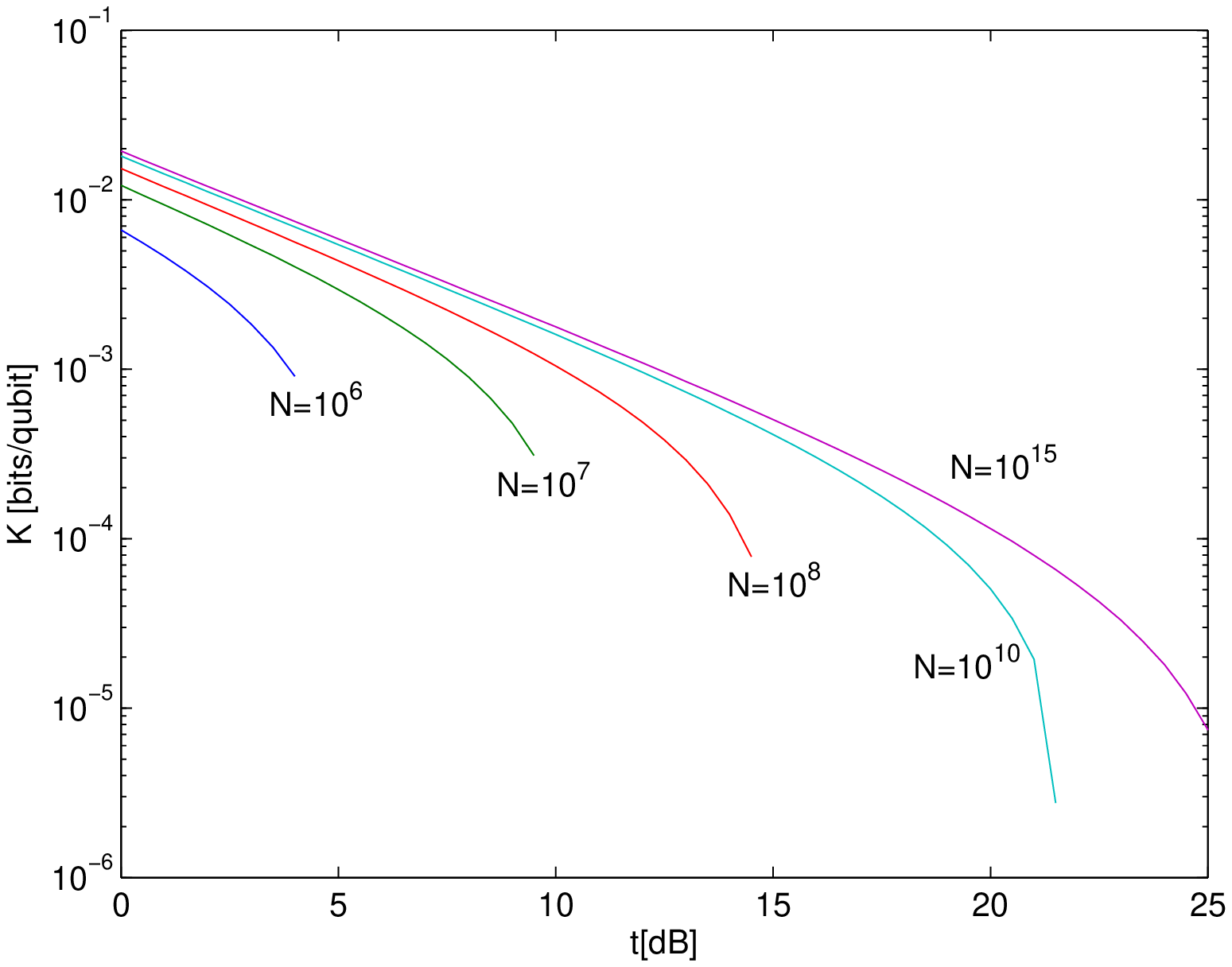}\\
\includegraphics[scale=0.6]{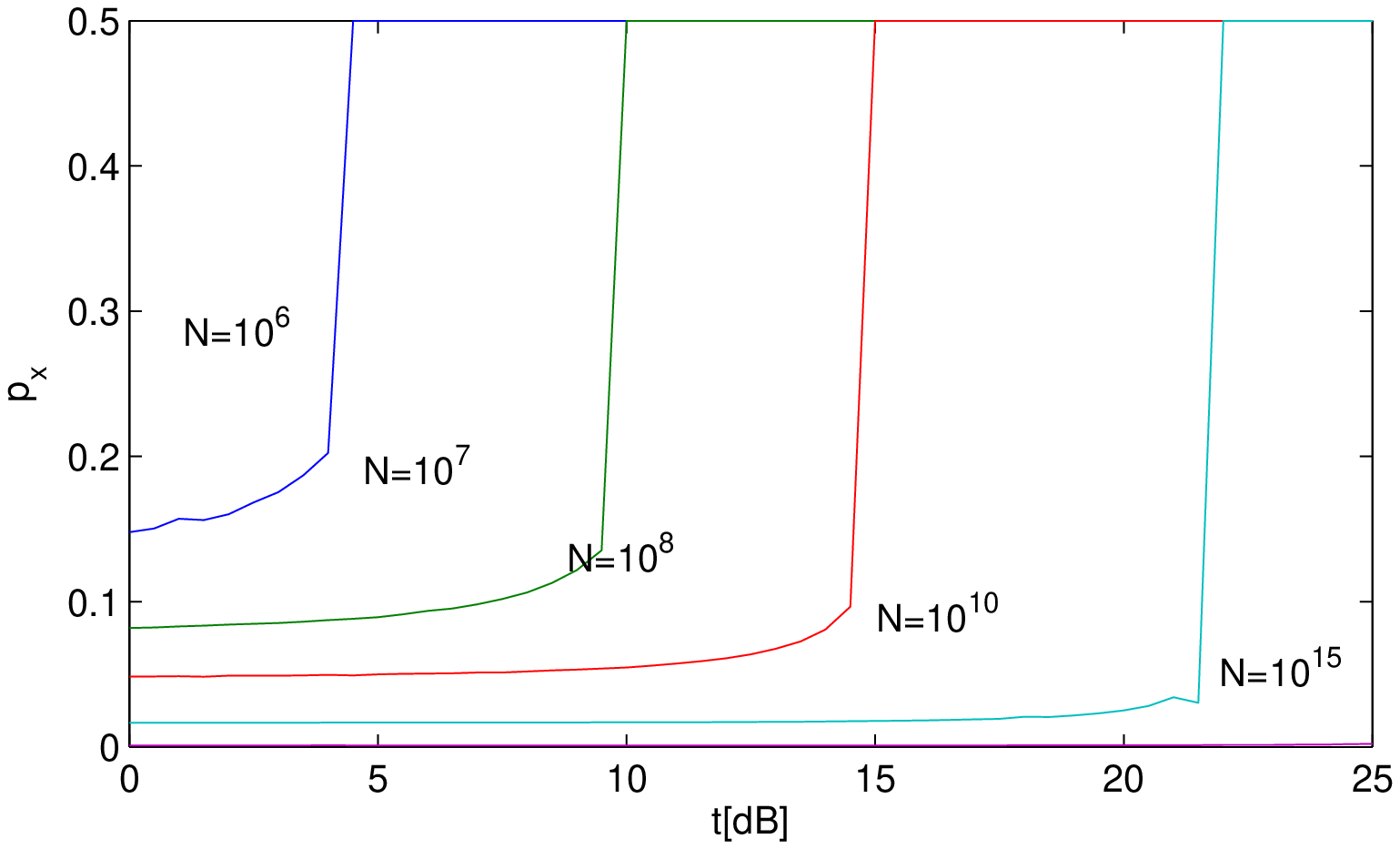}\\
\includegraphics[scale=0.6]{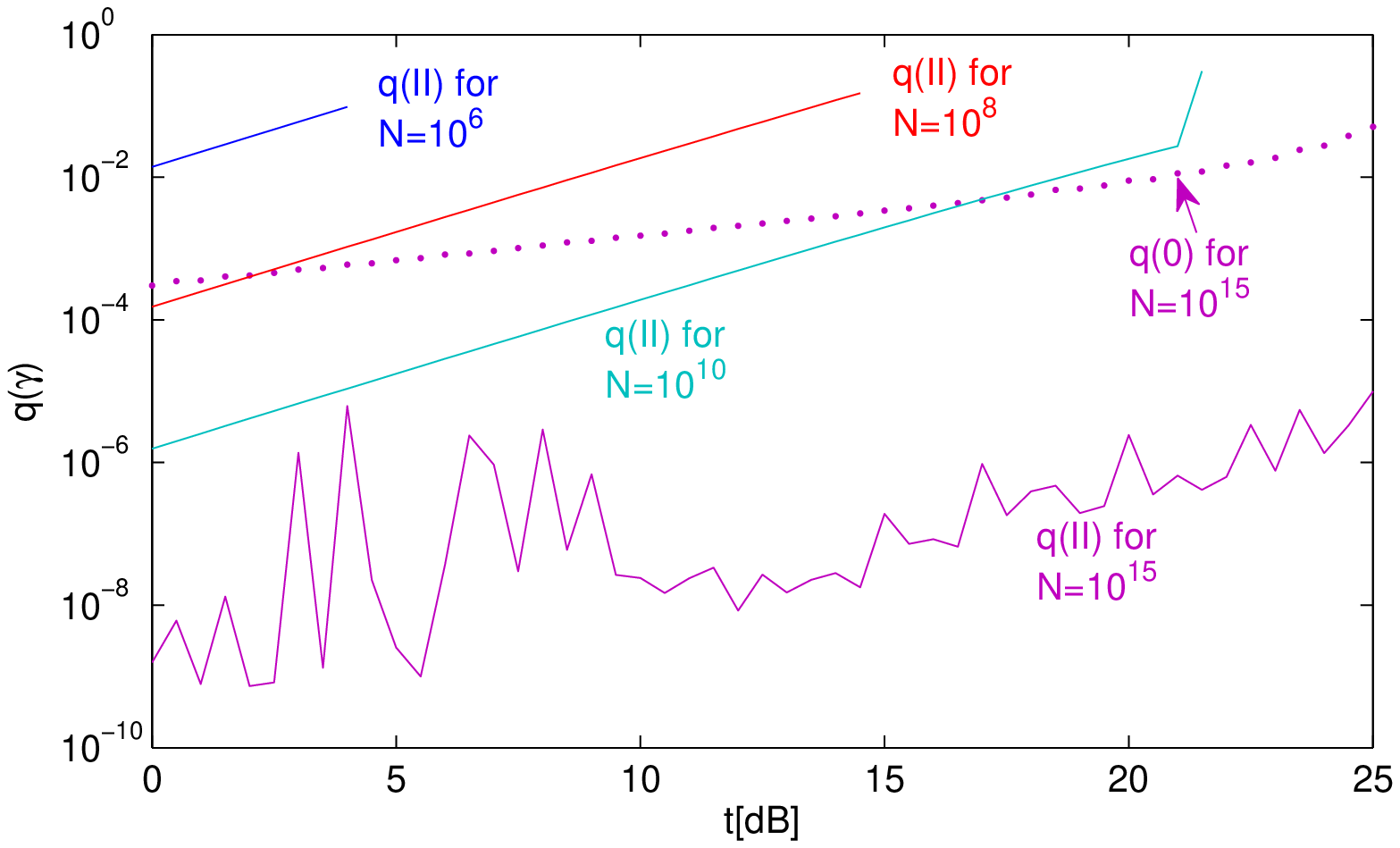}
\caption{Finite-key study of implementations of BB84 with weak coherent pulses for the three-intensity decoy state protocol described in the text, and assuming that only the intensity $\mu_{\mathrm{I}}$ is used for the key. As a function of the transmittivity of the channel $t$: upper graph, secret key rate $K$ from eq.~(\ref{boundwith1}); middle graph: corresponding optimal values of $p_X$; lower graph: corresponding values of $q_{\mathrm{\emptyset}}$ and $q_{\mathrm{II}}$ (regarding the large fluctuations in $q_{\mathrm{II}}$ for $N=10^{15}$: we have not tried to optimize with further precision, given that the value is anyway $q_{\mathrm{II}}\sim 10^{-7}$). Parameters as in Fig.~\ref{figwithout}: $\eps=10^{-5}$, $\eps_{\EC}=10^{-10}$, $\mathrm{leak}_{\EC}(e)=1.05\,h(e)$, $Q=0.5\%$, $\eta=0.1$, $p_d=10^{-5}$.}\label{figwith1}
\end{figure}

We turn now to implementations with decoy states. As we said, we consider the case where the key is extracted only out of the signals of intensity $\mu_{\mathrm{I}}<\mu_{\mathrm{II}}$. In this case, Alice can set $p_X(\mathrm{II})=1$: whenever she sends out a pulse of intensity $\mu_{\mathrm{II}}$, she can prepare it in the $X$ basis because these pulses will anyway be used only for parameter estimation. Bob's value of $p_X$ of course cannot depend on the intensities, and is supposed to be the same as the $p_X(\mathrm{I})$. The bound to be optimized reads therefore
\ba
K&=&q_\mathrm{I}\,R(\mu_{I})\,p_Z(\mathrm{I})^2\,\left[S_\xi(A|E,\mathrm{I})-\Delta(n)-\mathrm{leak}_{\EC}(e_Z(\mathrm{I}))\right] \label{boundwith1}
\ea
where $S_\xi(A|E,\mathrm{I})=Y_0^L(\mathrm{I})+Y_1^L(\mathrm{I})\left[1-h\big(e^U_X(1)\big)\right]$ with the expressions (\ref{y0}), (\ref{y1}) and (\ref{ex1}). There is a new set of parameters that needs to be optimized, namely the probabilities $q_\gamma$ of using each intensity. The results are plotted in Fig.~\ref{figwith1}. We observe that, as expected, the rates are much better than the ones obtained without decoy states. The optimal rates can actually be achieved by several pairs of $(\mu_{\mathrm{I}},\mu_{\mathrm{II}})$; we fixed $\mu_{\mathrm{II}}=0.65$ and further optimized $\mu_{\mathrm{I}}$: we found that $\mu_{\mathrm{I}}\approx 0.5$, independent on $t$ and slightly depending on $N$. Again, far from the critical distance $p_X$ varies as $N^{-1/4}$. More interesting is the behavior of the $q_\gamma$: $q_\mathrm{II}$ decreases with $N$, as expected; $q_\mathrm{\emptyset}$ however is non-zero only for $N=10^{15}$. This behavior can be easily understood because the only role of the zero-intensity pulses is to provide an estimate of the dark counts. Now, on the one hand the dark count rate is small, so one needs many signals to estimate it conveniently; on the other hand, the benefit of subtracting the dark count contribution is rather small.

Finally, we compare our results with previous estimates available in the literature. The very first papers on decoy states realized the importance of taking statistical fluctuations of the parameters into account  \cite{Wang05,Ma05,wan05}. These works differ from ours, in that they assume normal distribution for the fluctuations  (see \ref{fkgeneral}); moreover, they do not have the finite key correction $\Delta(n)$ and are therefore, strictly speaking, not providing lower bounds (neither were they claiming it, of course). However, their final estimates ultimately agree very well with ours. For instance, they had estimated that $N\approx 10^9-10^{10}$ is a ``reasonable number of signals'' and we arrive close to the asymptotic bound for similar values. More specifically, our plots for the achievable secret key rate are in remarkable agreement with those obtained in \cite{Ma05}, once some differences in the choice of the numerical values of parameters are taken into account. Of course, due to the different way fluctuations are introduced, some details differ. For instance, Ma and coworkers \cite{Ma05} found the optimal value of $q_\mathrm{\emptyset}$ to be approximately $4\times 10^{-2}$ already at $N=10^{10}$, while, as stressed just above, this value is zero in our approach for the same $N$. However, the discrepancy seems to be restricted to the choice of \textit{optimal} values for quantities that are anyway small; whence a suboptimal choice does not have a significant influence on the total result.

More recently, Hayashi and coworkers have provided another approach to compute a lower bound for decoy state protocols. When compared to ours, a striking fact is that they obtain a non-negligible finite key rate for $N$ as small as $10^4$ \cite{hay3}, while we do not obtain any key for $N<10^6$ signals. The comparison is not straightforward, since they are considering another decoy state protocol and the values of the parameters are different; nevertheless, their results suggests that our bounds might be improved.

\section{Entanglement-based implementations}
\label{sec4}

\subsection{Asymptotic bounds}

At the moment of writing, two asymptotic bounds are available for unconditional security of an entanglement-based implementation of the BB84 coding (BBM92 protocol). Under the squashing model for Bob's detectors, whose validity has been proved for BB84 coding \cite{bea08,tsu08}, Ma, Fung and Lo \cite{ma07} proved
\ba
S(A|E)&=&1-h(\mathbf{e_X})\,.\label{ebasymp1}
\ea
This means that, even if the source is not a single-pair source, all its imperfections are taken into account in the measured error rate, a feature anticipated by Koashi and Preskill \cite{koa03}. This result is remarkable, since it is formally identical to the one obtained for single-photon sources. As such, for the finite key-bound within our formalism we can refer to Ref.~\cite{scaren1}.

More recently, Koashi and coworkers have proved a different bound \cite{koa08}, which differs in the treatment of double-click events. In squashing, a physical double-click event is taken into account by adding a random bit to the raw key; the fraction of such events does not need to be measured. In the present approach, the double-click events are deleted from the raw key but their fraction $\delta_{2c}$ is estimated. Let $\mathbf{R}$ be the detection rate including double clicks, which is also the detection rate in the squashing model; and let $\mathbf{R'}$ the rate obtained once double-click events are removed (i.e. $\mathbf{R}-\mathbf{R'}$ is the measured number of double clicks). Asymptotically one has the exact estimate
\ba
\delta_{2c}&=&\frac{\mathbf{R}-\mathbf{R'}}{\mathbf{R}}\,.
\ea
The error rates observed in the raw key for the present approach are written $\mathbf{e'_X}$ and $\mathbf{e'_Z}$; they are related to the error rates that would be obtained by processing the same data with the squashing model through
\ba\mathbf{e_{X,Z}}&=&(1-\delta_{2c})\,\mathbf{e'_{X,Z}} +\delta_{2c}/2\,.\label{eeprime}\ea
In particular, in the case where the $\mathbf{e'_{X,Z}}$ are very small (e.g. for very high optical visibility), the present approach shows basically no errors. Specifically, let $F(\delta_{2c})\equiv (1-4\delta_{2c})/(1-\delta_{2c})$: for $\mathbf{e'_X}\lesssim 0.08\,F(\delta_{2c})$ one has\footnote{At this stage, it is useful to explain some difference in notation between us and Ref.~\cite{koa08}. Our $\mathbf{e'_X}$ and $\mathbf{e'_Z}$ are the error rates in the raw key, i.e. with the double-click events already removed; Koashi and co-workers assume $\mathbf{e'_X}=\mathbf{e'_Z}=\frac{\epsilon}{1-\delta}$. Our expression (\ref{saekoashi}) is obtained by inserting eq.~(20) into $1-\tau(\delta,\epsilon)/(1-\delta)$ from eq.~(3). Indeed, in our case $S(A|E)$ is Eve's uncertainty per bit of the raw key; the global factor $(1-\delta)$ will be accounted for in the detection rate $\mathbf{R'}$ defined below.}
\ba
S(A|E)&=&F(\delta_{2c})\,\left[1-h\left(\frac{\mathbf{e'_{X}}}{F(\delta_{2c})}\right)\right]\,.
\label{saekoashi}
\ea 
Indeed, in the regime of small errors, the asymptotic secret key rate $K$ computed with (\ref{saekoashi}) is larger than the one computed from (\ref{ebasymp1}). However, the former implies the estimation of an additional parameter, namely $\delta_{2c}$. It is therefore interesting to compare the two approaches in the finite-key scenario.

\subsection{Finite-key security bounds and \textit{a priori} expected values}

The finite-key secret-key rate associated to the first approach (\ref{ebasymp1}) is
\ba
K&=&\mathbf{R}\,p_Z^2\,\left[1-h(e_X^U)-\Delta(n)-\mathrm{leak}_{\EC}(\mathbf{e_Z})\right] \label{boundsquash}
\ea
with $e_X^U=\mathbf{e_X}+\xi(m,2)$. As in the case of single-photon sources, the only parameter that needs to be estimated is the error rate (so $n_{PE}=1$). Similarly, for the second approach (\ref{saekoashi}) one obtains
\ba
K&=&\mathbf{R'}\,p_Z^2\,\left\{F(\delta^U_{2c})\left[1-\,h\left(\frac{{e'}_X^U}{F(\delta^U_{2c})}\right)\right]-\Delta(n)-\mathrm{leak}_{\EC}(\mathbf{e'_Z})\right\}\label{bounddoubleclick}
\ea
with ${e'}_X^U=\mathbf{e'_X}+\xi(m,2)$ and $\delta^U_{2c}=(\mathbf{R}-\mathbf{R'})/(\mathbf{R})+\xi(N,2)$. Obviously here $n_{PE}=2$. 

In order to compare the two approaches \textit{a priori}, we need to insert an expected value of the measured parameters and run the optimization over the free parameters left. We consider an implementation with continuous-wave pumping, following paragraph VII.A.1 of \cite{sca08}, where all details can be found; for a more detailed description, see \cite{ma07}, especially eqs (9) and (10). The pump intensity is such that $\mu'$ pairs are produced within the coincidence window $\Delta\tau$; we work in the limit $y\equiv \mu'\Delta\tau\ll 1$ and neglect dark counts on Alice's side. Therefore, whenever Alice detects a photon, which happens with probability $\approx y$, the signal traveling to Bob is distributed according to $p_A(1)\approx 1$, $p_A(2)\approx y$ and $p_A(n>2)\approx 0$. The expected values for the single-click rate $R_{1c}$ and the corresponding error rate $Q$ are given by
\ba
R_{1c}/y&=&R_p/y+R_d/y\,\approx\,t\eta\left[p_A(1)+p_A(2)(2-t\eta)\right]+2p_d\left[p_A(1)(1-t\eta) + p_A(2)(1-t\eta)^2\right]\,,\\
Q&=&[(1-V+y)R_p+R_d]/2R_{1c}
\ea
(note the presence of the two-pair fraction $y$ as a linear decrease in the observed two-photon visibility $V$). The detection rate of double clicks is
\ba
R_{2c}/y&=&p_A(2)\demi(t\eta)^2\,+\,[p_A(1)+p_A(2)(1-t\eta)][t\eta p_d\,+\,(1-t\eta)p_d^2]\,.
\ea
So we have the \textit{a priori} expected values $R=R_{1c}+R_{2c}$, $R'=R_{1c}$ and $\delta_{2c}=R_{2c}/(R_{1c}+R_{2c})$. As for the error rates, we identify $e'_X=e'_Z=Q$, whence (\ref{eeprime}) implies $e_X=e_Z=(1-\delta_{2c})\,Q +\delta_{2c}/2$.

\begin{figure}[ht]
\includegraphics[scale=0.8]{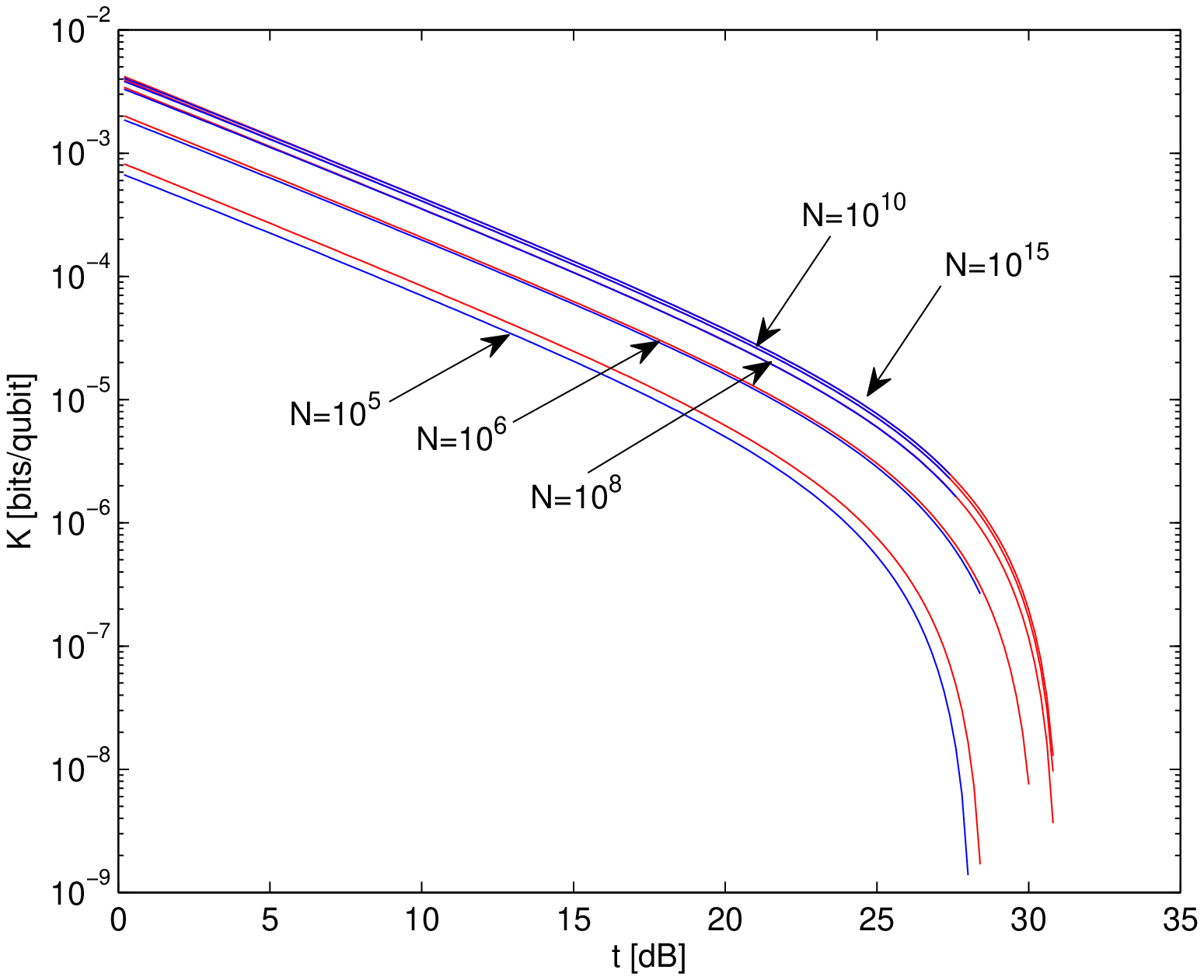}
\caption{Key rate $K$ as a function of the attenuation $t$ for entanglement-based implementations of the BB84 coding. Red curves: bound with squashing (\ref{boundsquash}), adapted from the asymptotic bound of Ref.~\cite{ma07}. Blue curves: bound with estimate of double-clicks (\ref{bounddoubleclick}), adapted from the asymptotic bound of Ref.~\cite{koa08}. Parameters as in Figs \ref{figwithout} and \ref{figwith1}: $\eps=10^{-5}$, $\eps_{\EC}=10^{-10}$, $\mathrm{leak}_{\EC}(e)=1.05\,h(e)$, $V=0.99$ (corresponding to $Q=0.5\%$ if one neglects the effect of double pairs), $\eta=0.1$, $p_d=10^{-5}$.}\label{figebrates}
\end{figure}

The result of the numerical optimization over $y$ and the finite-key parameters is shown in Fig.~\ref{figebrates}. As expected, for small number of signals the squashing bound outperforms the double-click one, because the latter needs to estimate a second parameter. For larger number of signals, the two bounds give identical rates (the very small difference can be attributed to our approximations, like neglecting the cases when $n>2$ pairs are created). The values of $y$ and $p_X$ are also basically identical for both bounds. As observed in the prepare-and-measure schemes, $y$ varies little with $N$ ($y\approx 0.05$ for $N=10^5$, $y\approx 0.1$ for large $N$), while $p_X$ scales as $\sim N^{-1/4}$.

\section{Conclusion}

In summary, we have provided security bounds for keys of finite length for several practical implementations of the BB84 coding. The bounds for prepare-and-measure implementations without decoy states and for entanglement-based implementations guarantee unconditional security; the bound for prepare-and-measure implementations with decoy states has been derived using a simplified treatment of the statistical fluctuations.

We have computed these bounds for \textit{a priori} expected values of the parameters that will be observed, thus providing some guidelines for the design of experiments. In all cases, for $N\gtrsim 10^{15}$, we recover the asymptotic bounds (compare e.g. with the plots in \cite{sca08}). However, prepare-and-measure implementations based on weak coherent pulses seem to require at least $N\sim 10^7$ signals to produce a key; while implementations using entangled states, similarly to the ideal single-photon case, provide a key already for $N\sim 10^5$.

Let us conclude by a critical review of the possible extensions and open issues. The bounds presented in this paper have been derived under some assumptions. Some of them are assumptions on Alice and Bob, mostly inherited from the asymptotic studies from which $S(A|E)$ was obtained. Specifically: 
\begin{itemize}

\item First, we recall that, in the case of decoy states, we have used a partial treatment of the statistical fluctuations; also, we have provided an actual bound only for a specific choice (one intensity for the key signals, two for the decoys, one of which being zero).

\item In all weak coherent pulses implementations we have supposed that there is no phase coherence between successive pulses; in the case of entanglement-based schemes, we have assumed continuous pumping.

\item All the bounds we used assume that the bit values `0' and `1' appear the same number of times in both Alice's and Bob's raw keys. A systematic deviation from this assumption is expected if the detectors have different efficiencies, which is often the case in practice. The tools to study this case are available in the asymptotic scenario \cite{fung08}, their finite-key generalization should be the object of further work. Of course, in case one bit value is more frequent than the other, a conservative security bound is obtained by adding the number of excess bits to the information of Eve to be removed during privacy amplification; therefore one can use our formulas with this modification.

\item The prepare-and-measure bounds given above are not valid for Plug-and-Play configurations, even if the difference is ultimately expected to be small. The reason is that the ``source'' on Alice's side cannot be assumed to produce exact weak coherent pulses, because these pulses are obtained by attenuating an in principle unknown strong incoming signal. An asymptotic bound for unconditional security of Plug-and-Play configurations has been given in Ref.~\cite{zha08}. Its generalization to finite keys may be done by following the same procedure as in this paper.

\item When we provide \textit{a priori} expected values, we have always performed an optimization over $p_X$. Some systems may be such that this optimization cannot be easily performed (e.g., in a passive detection setup, one would have to change the beam-splitter that chooses between the bases).

\end{itemize}

A second group of assumptions is related to the fact that our bounds may be the object of improvements:
\begin{itemize}

\item First of all, the fact of having used the formalism developed in \cite{scaren1,scaren2} guarantees unconditional security, but it is not known whether the bounds are tight. Indeed, all the different approaches to security are known to coincide in the asymptotic regime, but this is not yet clear for the finite-key regime --- and we hinted in \ref{secdecoy} to an actual discrepancy between ours and other estimates in the case of decoy states implementations. Most of the information-theoretical estimates are generally regarded to be tight \cite{rennerthesis}; however, we have bounded statistical fluctuations using absolute errors (\ref{eqpe}); improvements may be obtained by using relative errors.

\item We have computed the security bounds for the case when the extraction of the secret key is done through one-way post-processing without pre-processing. In principle, the tools are available to compute finite-key bounds for two-way post-processing and including pre-processing \cite{scaren2}. For typical error rates, the improvements are supposed to be significant only close to the critical distance.

\item For simplicity, we have considered asymmetric implementations of the BB84 coding, in which the $Z$ basis is used for the key and the $X$ basis for parameter estimation. If both bases are used for the key (while each basis serving to estimate Eve's attack on the other), one obtains similar more complicated expressions, but basically (assuming $p_X\leq p_Z$) the effect is to increase $K$ by a factor $1+(p_X/p_Z)^2$. A similar argument can be made in the case of decoy states protocols, where we have assumed for simplicity that only one intensity is used for the key.

\end{itemize}

\section*{Acknowledgment}
We thank all the participants to the workshop ``Quantum cryptography with finite resources'' (Singapore, 4-6 December 2008) for very valuable comments. We are grateful to Hongwei Li (USTC, Hefei, China) for bringing to our attention the possibility of improving the estimate given in Eq.~(\ref{eqpe}). This work was supported by the National Research Foundation and the Ministry of Education, Singapore.

\newpage

\newpage

\newpage

\begin{thebibliography}{19}
\expandafter\ifx\csname natexlab\endcsname\relax\def\natexlab#1{#1}\fi
\expandafter\ifx\csname bibnamefont\endcsname\relax
  \def\bibnamefont#1{#1}\fi
\expandafter\ifx\csname bibfnamefont\endcsname\relax
  \def\bibfnamefont#1{#1}\fi
\expandafter\ifx\csname citenamefont\endcsname\relax
  \def\citenamefont#1{#1}\fi
\expandafter\ifx\csname url\endcsname\relax
  \def\url#1{\texttt{#1}}\fi
\expandafter\ifx\csname urlprefix\endcsname\relax\def\urlprefix{URL }\fi
\providecommand{\bibinfo}[2]{#2}
\providecommand{\eprint}[2][]{\url{#2}}

\bibitem{bb84}
\bibinfo{author}{\bibfnamefont{C.}~\bibnamefont{Bennett}} \bibnamefont{and}
  \bibinfo{author}{\bibfnamefont{G.}~\bibnamefont{Brassard}}, in
  \emph{\bibinfo{booktitle}{Proceedings of IEEE International Conference on
  Computers, Systems and Signal Processing, Bangalore, India, December 1984,
  pp. 175 - 179.}} (\bibinfo{year}{1984}).

\bibitem{eke91} A.K. Ekert, Phys. Rev. Lett. \textbf{67}, 661 (1991)

\bibitem{gis02} N. Gisin, G. Ribordy, W. Tittel, H.
Zbinden, Rev. Mod. Phys. \textbf{\bibinfo{volume}{74}},
\bibinfo{pages}{145} (2002)

\bibitem{dus06} M. Du\v{s}ek, N. L\"utkenhaus, M. Hendrych, Progress in Optics \textbf{\bibinfo{volume}{49}}, Edt. E. Wolf (Elsevier), \bibinfo{pages}{381} (2006)

\bibitem{sca08} V. Scarani, H. Bechmann-Pasquinucci, N.J. Cerf, M. Du\v{s}ek, N. L\"utkenhaus, M. Peev, arXiv:0802.4155

\bibitem{lo08} H.-K. Lo, Y. Zhao, arXiv:0803.2507

\bibitem{may96} D. Mayers, in: \emph{\bibinfo{booktitle}{Advances in Cryptology \,---\, Proceedings of Crypto '96}} (\bibinfo{publisher}{Springer Verlag, Berlin}), p. 343 (1996).

\bibitem{LoChau} H.-K. Lo, H. F. Chau, Science \textbf{283}, 2050 (1999)

\bibitem{ShoPre00} P.W. Shor, J. Preskill, Phys. Rev. Lett. \textbf{85}, 441 (2000).

\bibitem{Mayers01} D. Mayers, Journal of the ACM \textbf{48}, 351 (2001); and quant-ph/9802025.

\bibitem{BenOr02} M. Ben-Or, Security of {BB84} {QKD} Protocol, Slides available at \url{http://www.msri.org/publications/ln/msri/2002/quantumintro/ben-or/2/}

\bibitem{KGR} B. Kraus, N. Gisin, R. Renner, Phys. Rev. Lett. {\bf 95}, 080501 (2005); R. Renner, N. Gisin, B. Kraus, Phys. Rev. A {\bf 72}, 012332 (2005).

\bibitem{rennerthesis} R. Renner, \textit{Security of Quantum Key Distribution}, PhD thesis, Diss.~ETH No 16242; published in: Int. J. Quant. Inf. \textbf{6}, 1 (2008)

\bibitem{Koashi} M. Koashi, J. of Phys. Conference Series \textbf{36}, 98 (2006)

\bibitem{kur01} C. Kurtsiefer, P. Zarda, S. Mayer, H. Weinfurter, J. Mod. Opt. \textbf{48}, 2039 (2001).

\bibitem{makarov} V. Makarov, D. R. Hjelme, J. Mod. Opt. \textbf{52},691 (2005); V. Makarov, A. Anisimov, J. Skaar, Phys. Rev. A \textbf{74}, 022313 (2006)

\bibitem{zhao07} Y. Zhao, C.-H. F. Fung, B. Qi, C. Chen, H.-K. Lo, Phys. Rev. A \textbf{78}, 042333 (2008)

\bibitem{aci07}A. Ac\'{\i}n, N. Brunner, N. Gisin, S. Massar, S. Pironio, V. Scarani, Phys. Rev. Lett. \textbf{98}, 230501 (2007)

\bibitem{pir09} S. Pironio, A. Ac\'{\i}n, N. Brunner, N. Gisin, S. Massar, V. Scarani, New J. Phys. \textbf{11}, 045021 (2009)

\bibitem{ina07} H. Inamori, N. L\"utkenhaus, D. Mayers, Eur. J. Phys. D \textbf{41}, 599 (2007), and quant-ph/0107017.

\bibitem{BHLMO05} M. Ben-Or, M. Horodecki, D.W. Leung, D. Mayers, J. Oppenheim, in
  \emph{\bibinfo{booktitle}{Second Theory of Cryptography Conference {TCC}}}
  (\bibinfo{publisher}{Springer}, \bibinfo{year}{2005}), vol.
  \bibinfo{volume}{3378} of \emph{\bibinfo{series}{Lecture Notes in Computer
  Science}}, pp. \bibinfo{pages}{386--406}, \bibinfo{note}{and quant-ph/0409078}.

\bibitem{KRBM07}
\bibinfo{author}{\bibfnamefont{R.}~\bibnamefont{K\"onig}},
  \bibinfo{author}{\bibfnamefont{R.}~\bibnamefont{Renner}},
  \bibinfo{author}{\bibfnamefont{A.}~\bibnamefont{Bariska}},
  \bibinfo{author}{\bibfnamefont{U.}~\bibnamefont{Maurer}},
  \bibinfo{journal}{Phys.\ Rev.\ Lett.} \textbf{\bibinfo{volume}{98}},
  \bibinfo{pages}{140502,} (\bibinfo{year}{2007}).

\bibitem{LoChauArdehali} \bibinfo{author}{\bibnamefont{H.-K. Lo, H. F. Chau, M. Ardehali}},
\bibinfo{journal}{J. Cryptology} \textbf{\bibinfo{volume}{18}},
\bibinfo{pages}{133} (2005); and \eprint{quant-ph/9803007}.

\bibitem{Ma05} X. Ma, B. Qi, Y. Zhao, H.-K. Lo, Phys. Rev. A \textbf{72}, 012326 (2005). 

\bibitem{Wang05} X.-B. Wang, Phys. Rev. Lett. \textbf{94}, 230503 (2005). 

\bibitem{mey06} T. Meyer, H. Kampermann, M. Kleinmann, D. Bru\ss, Phys. Rev. A \textbf{74},
042340 (2006).

\bibitem{hay2} M. Hayashi, Phys. Rev. A \textbf{76}, 012329 (2007).

\bibitem{hase} J. Hasegawa, M. Hayashi, T. Hiroshima, A. Tanaka, A. Tomita, arXiv:0705.3081.

\bibitem{scaren1} V. Scarani, R. Renner, Phys. Rev. Lett. \textbf{100}, 200501 (2008)

\bibitem{scaren2} V. Scarani, R. Renner, in: \textit{Proceedings of TQC2008}, Lecture Notes in Computer Science \textbf{5106} (Springer Verlag, Berlin), pp. 83-95 (2008); and arXiv:0806.0120

\bibitem{hay3} J. Hasegawa, M. Hayashi, T. Hiroshima, A. Tomita, arXiv:0707.3541.

\bibitem{bbm92}C.H. Bennett, G. Brassard, N.D. Mermin, \bibinfo{journal}{Phys. Rev. Lett.} \textbf{\bibinfo{volume}{68}},
\bibinfo{pages}{557} (1992)

\bibitem{DevWin05}
\bibinfo{author}{\bibfnamefont{I.}~\bibnamefont{Devetak}},
  \bibinfo{author}{\bibfnamefont{A.}~\bibnamefont{Winter}},
  \bibinfo{journal}{Proc.\ R.\ Soc.\ Lond. A} \textbf{\bibinfo{volume}{461}},
  \bibinfo{pages}{207} (\bibinfo{year}{2005}).

\bibitem{hay4} M. Hayashi, Phys. Rev. A \textbf{79}, 032303 (2009)

\bibitem{gotlo} D. Gottesman, H.-K. Lo, IEEE Trans. Inf. Theory \textbf{49}, 457 (2003).

\bibitem{chri08} M. Christandl, R. K\"onig, R. Renner, Phys. Rev. Lett. \textbf{102}, 020504 (2009) 


\bibitem{cover} T.M. Cover, J.A. Thomas, \textit{Elements of Information Theory}, Wiley Series in Telecommunications (Wiley, New York, 1991); we refer to Theorem 12.2.1 and Lemma 12.6.1.

\bibitem{lut99} N. L\"{u}tkenhaus, Phys. Rev. A \textbf{\bibinfo{volume}{59}},
\bibinfo{pages}{3301} (1999)


\bibitem{lopre} H.-K. Lo, J. Preskill, \bibinfo{journal}{Quant. Inf. Comput.} \textbf{\bibinfo{volume}{8}},
\bibinfo{pages}{431} (2007)

\bibitem{gllp} D. Gottesman, H.-K. Lo, N. L\"utkenhaus, J. Preskill, Quant. Inf. Comput. \textbf{\bibinfo{volume}{4}},
\bibinfo{pages}{325} (2004)

\bibitem{fun06a} C.-H.F. Fung, K. Tamaki, H.-K. Lo, Phys. Rev. A \textbf{\bibinfo{volume}{73}}, \bibinfo{pages}{012337} (2006)

\bibitem{kra07} B. Kraus, C. Branciard, R. Renner, Phys. Rev. A \textbf{\bibinfo{volume}{75}},
\bibinfo{pages}{012316} (2007)

\bibitem{bea08} N.J. Beaudry, T. Moroder, N. L\"utkenhaus, Phys. Rev. Lett. \textbf{\bibinfo{volume}{101}},
\bibinfo{pages}{093601} (2008)

\bibitem{tsu08} T. Tsurumaru, K. Tamaki, Phys. Rev. A \textbf{\bibinfo{volume}{78}},
\bibinfo{pages}{032302} (2008)

\bibitem{mak08} V. Makarov, A. Anisimov, S. Sauge, arXiv:0808.3408

\bibitem{hwa03} W.-Y. Hwang, Phys. Rev. Lett. \textbf{\bibinfo{volume}{91}},
\bibinfo{pages}{057901} (2003)

\bibitem{wan05} X.-B. Wang, Phys. Rev. Lett. \textbf{\bibinfo{volume}{94}},
\bibinfo{pages}{230503}(2005)

\bibitem{lo05}H.-K. Lo, X. Ma, K. Chen, Phys. Rev. Lett. \textbf{\bibinfo{volume}{94}},
\bibinfo{pages}{230504} (2005)

\bibitem{ma07} X. Ma, C.-H.\,F.~Fung, H.-K.~Lo,
\bibinfo{journal}{Phys. Rev. A} \textbf{\bibinfo{volume}{76}},
\bibinfo{pages}{012307} (2007).


\bibitem{koa03} M. Koashi, J.~Preskill, Phys. Rev. Lett. \textbf{\bibinfo{volume}{90}},
\bibinfo{pages}{057902} (2003)

\bibitem{koa08} M. Koashi, Y. Adachi, T. Yamamoto, N. Imoto,
\eprint{arXiv:0804.0891}.

\bibitem{fung08} C.-H.F. Fung, K. Tamaki, B. Qi, H.-K. Lo, X. Ma, Quantum Inf. Comput. \textbf{9}, 131 (2009) 

\bibitem{zha08} Y. Zhao, B. Qi, H.-K. Lo, Phys. Rev. A \textbf{\bibinfo{volume}{77}},
\bibinfo{pages}{052327} (2008)

\end{thebibliography}
\end{document}